\documentclass[showpacs,preprintnumbers,amsmath,amssymb]{revtex4}

\usepackage[dvips]{graphicx}
\usepackage{amsmath}

\newcommand{\CO}{{\cal O}}

\newcommand{\vecs}[1]{\mbox{\boldmath${#1}$}}
\newcommand{\vect}[1]{\mbox{\boldmath\tiny${#1}$}}

\begin{document}

\title{Affleck Dine leptogenesis via multiple flat directions}

\author{Kohei Kamada$^1,^2$ and Jun'ichi Yokoyama$^2$} 
\affiliation{$^1$Department of Physics, Graduate School of Science,  
The University of Tokyo, Tokyo 113-0033, Japan\\
$^2$Research Center for the Early Universe (RESCEU),
Graduate School of Science, The University of Tokyo,Tokyo 113-0033, Japan
}


\date{\today}

\begin{abstract}
We investigate the Affleck-Dine mechanism when multiple flat directions have large values simultaneously. 
We consider in detail the case when both $LH_u$ and $H_uH_d$ flat directions are operative with a non-renormalizable superpotential. 
In case Hubble induced A-terms are present for these two flat directions, their initial values are determined completely by the potential and there are no ambiguities how they are mixed. 
Moreover, CP is violated even when the Hubble parameter is large due to the Hubble induced A-term and cross coupling in F-term, so that the  lepton asymmetry is generated just after the end of inflation. 
As a result, compared with the case of single flat direction, the resultant lepton-to-entropy ratio is enhanced by a factor of $H_{osc}/m_{3/2}$, where $H_{osc}$ is the Hubble parameter at the onset of oscillation and $m_{3/2}$ is the gravitino mass.   
However, when Hubble induced A-terms do not exist, there remains indefiniteness of initial phases and CP is violated spontaneously by the phase difference between initial phase and potential minima of the hidden-sector induced A-terms. 
Therefore, CP-violation is not effective until the onset of the oscillation of scalar fields around the origin and there is suppression factor from thermal effect as is the case of single flat direction. 
In this case, the amplitude of baryon isocurvature perturbation imposes constraints on the model parameters. 
\end{abstract}

\maketitle

\section{Introduction}
The origin of the observed baryon asymmetry of the Universe\cite{Spergel:2006hy}
is one of the most important problems in both cosmology and particle physics which cannot be explained within the standard model of particle physics (SM) \cite{Sakharov:1967dj}. 
Inflation in the early universe \cite{Sato:1980yn}, which is required in order to solve many cosmological problems such as the flatness and the horizon problems and to account for the origin of primordial fluctuations, washes out any preexist baryon asymmetry, so baryogenesis \cite{Kuzmin:1970nx,Yoshimura:1978ex}
must take place after inflation but before the big bang nucleosynthesis (BBN) \cite{Olive:1999ij} epoch. 

As for the physics beyond SM, one of the best motivated models is supersymmetry (SUSY) \cite{Nilles:1983ge}, because it can solve the hierarchy problem and realizes the gauge coupling unification. 
Generic supersymmetric models have flat directions in field space where scalar potential vanishes at the classical level and their dynamics is very important in the broad context. 
The Affleck-Dine (AD) mechanism \cite{Affleck:1984fy,Linde:1985gh,Dine:1995kz} of baryo(lepto)genesis makes use of scalar fields having a large amplitude along the flat directions. 
In this scenario, the angular momentum, or the rotation around the origin, of scalar fields  that carry baryon or lepton number represents the number density of the field. 
Note that lepton asymmetry can be converted to baryon asymmetry through the sphaleron effect \cite{Kuzmin:1987wn}, which violates $B+L$ at the electroweak scale, where $B$ and $L$ are baryon and lepton charges, respectively. 
Therefore the source of baryon asymmetry in this scenario is large values of the scalar fields as the initial conditions and the phase difference between the potential minimum and the initial condition, which represents CP-violation. 
Later Dine et al.\cite{Dine:1995kz} analyzed behaviors of the field that parameterizes a flat direction (We call it the AD field.) during and after inflation taking non-renormalizable superpotential of the form 
\begin{equation}
W=\frac{\varphi^{n+3}}{M^n}
\end{equation}
and SUSY breaking in the early universe into account. 
Here $\varphi$ is the AD field and $M$ is some cut-off scale. 
In this scenario, SUSY breaking in the early universe induces both a possibly negative mass term of the form, $-cH^2|\varphi|^2$, and CP-violating A-terms of the form, $(aH/M)\varphi^{n+3}+\mathrm{h.c.}$ where $c$ (real) and $a$ (complex) are numerical factors of order unity and $H$ is the Hubble parameter. 
The negative mass term assures the flat direction to have a nonvanishing vacuum expectation value (vev) and the difference of phase between $a$ and the coupling constant of A-term from hidden sector of the form, $(Am_{3/2}/M)\varphi^{n+3}+{\rm h.c.}$, which is a consequence of SUSY breaking at low energy, provides CP-violation. 
When the Hubble induced A-terms do not exist, initial value of AD field in the angular direction is determined randomly during inflation and CP is violated spontaneously by the phase difference between initial phase of AD field and phase minima of the A-term from hidden sector. 
In this scenario, in both cases, the net baryon-to-entropy ratio is in proportion to the cut-off scale and the reheating temperature. 
Furthermore, it has been pointed out that thermal effect induces an early coherent oscillation of AD field and suppresses the resultant baryon asymmetry  \cite{Anisimov:2000wx,Allahverdi:2000zd,Asaka:2000nb,Fujii:2001zr}. 
 
Those analyses restricted only in the case where the configuration of flat direction can be parameterized in terms of one complex scalar field. 
There are, however, many flat directions even in the minimal supersymmetric standard model (MSSM) \cite{Gherghetta:1995dv},  some of which carry $B-L$ charge but others do not. 
Therefore it is very important which flat direction, if any, is selected as the AD field.  
Moreover, there are multiple flat directions which do not give rise to any F-term in the renormalizable limit\cite{Gherghetta:1995dv}. 
Such directions can get large values at the same time and we can no longer parameterize them by one scalar field. 
If some of them carry $B-L$ charge and others do not, the degree of the mixing of multiple flat directions  directly affects the net baryon asymmetry. 
It is not trivial whether the simple one-field analysis is applicable in such a case. 

The case where multiple flat directions have large values in the MSSM and its extensions have been considered in the literature \cite{Senami:2002kn,Senami:2001qn,Senami:2007wz,Enqvist:2003pb}. 
In this paper, we improve the evaluation of the potential and clarify how final lepton asymmetry is determined in the multifield context. 
In particular, we highlight the difference in the origin of CP-violation between the single-field case and the multiple-field case. 
We also study how much pre-inflationary information survives after inflation.

As mentioned above the general structure of the flat directions of the scalar potential in MSSM has been classified by Gherghetta, Kolda and Martin \cite{Gherghetta:1995dv}, but not all these flat directions remain flat in cosmological setting where scalar fields are expected to be distributed rather randomly \cite{Linde:1983gd}. 
For example, if Higgs fields have large values, only $L_iH_u (i=1,2,3)$  and $H_uH_d$ flat directions remain flat and  the other fields are driven to vanishingly small values. 
Here $L_i$ is left-handed slepton multiplet ($i$ is family index), $H_u$ is the Higgs multiplet that couples to $u$-type quarks and $H_d$ is the Higgs multiplet that couples to $d$-type quarks and leptons. 
In this paper, we concentrate on such a situation, which is similar to the case \cite{Senami:2002kn}. 
We investigate the potential for these scalar fields in detail and estimate the net lepton asymmetry. 
We find that Hubble induced A-terms play critical roles in both the survival of pre-inflationary information and the dynamics of AD fields. 
If Hubble induced A-terms exist, our result for the resultant lepton asymmetry is consistent with the case of \cite{Senami:2002kn}, but the dynamics of one of the scalar fields is slightly different. 
Moreover, if Hubble induced A-terms do not exist, there are almost no differences to the dynamics of scalar fields between single flat direction case and multiple flat directions case. 
In such a case, we must pay attention to the baryon isocurvature perturbation as is the case with Ref.\cite{Yokoyama:1993gb}

The structure of the paper is as follows. 
In the following section we present the relevant part of the MSSM in which we are interested. 
Then it is argued how CP-violating term is introduced and what is responsible for CP-violation. 
Furthermore we show how multiple flat directions are mixed in our model and how natural we consider the case where multiple flat directions have large value. 
In section \ref{Dyna}, AD mechanism in our model is discussed. 
The classical evolution along two flat directions is studied.  
In section \ref{noa}, we comment on the case when the Hubble induced A-terms do not exist. 
We also discuss the baryon isocurvature perturbation. 
Finally, section \ref{conc} is devoted to the conclusion of the dynamics of scalar fields and leptogenesis in our model.

\section{Model \label{sec-mod}}

\subsection{Potential \label{sec-pot}}
We adopt a non-renormalizable superpotential of the form,
\begin{equation}
\delta W=\frac{\lambda_{Lij}}{2M}(L_iH_u)(L_jH_u)+\frac{\lambda_H}{2M}(H_uH_d)(H_uH_d), \label{nonsup}
\end{equation}
in addition to the lepton sector of the renormalizable superpotential of MSSM \cite{Nilles:1983ge}, 
\begin{equation}
W_{\rm ren}={\overline e}_i {\bf y_e}_{ij} L_j H_d, 
\end{equation} 
where $\overline e$ is right-handed slepton and ${\bf y_e}$ is the Yukawa-coupling for lepton sector. 
Here $M$ is some cut-off scale, $\lambda_{Lij}$ and $\lambda_H$ are coupling constants and $i$ is family index. 
For example, if there is a right-handed Majorana neutrino multiplet whose mass $M$ is heavier than Hubble parameter, 
we can acquire the first term of \eqref{nonsup} by integrating out the right-handed sneutrino. 
We choose a basis which $\lambda_{Lij}$ is diagonal.
Note that this superpotential gives a neutrino mass when one of the Higgs fields gets a vev,
\begin{equation}
m_{\nu_{i}}=\frac{\lambda_{Lii}}{M}\langle H_u \rangle ^2=\frac{\lambda_{Lii}}{M} v^2 \sin^2 \beta. \label{neutrinomass}
\end{equation}
Here $v^2\equiv \langle H_u^0 \rangle^2+\langle H_d^0 \rangle^2\simeq 174$GeV and $\tan \beta \equiv \langle H_u^0 \rangle / \langle H_d^0 \rangle$. ($0$ means the neutral component of Higgs fields.)
To realize inflation, we introduce an inflaton sector besides the MSSM sector. 
Couplings between the inflaton sector and the MSSM sector arises only with the gravitational strength in the supergravity scalar potential, 
\begin{equation}
V=e^{K/M_G^2}\left(D_i W K^{i {\bar j}}D_{\bar j}W^* -\frac{3}{M_G^2}|W|^2\right), 
\end{equation}
and assuming non-minimal K${\ddot {\text a}}$hler potential, 
\begin{align}
\delta K_1&=\frac{a_a}{M_G^2}|I|^2|\phi_a|^2 \label{NMK1}\\
 \text{and} \ \ \delta K_2&=\frac{b_a}{M_G}I|\phi_a|^2+\rm{h.c.}, \label{NMK2}
\end{align}  
in addition to the canonical terms. 
Here $\phi_a$ is scalar fields of MSSM and $I$ is the inflaton, $a$ is a label to distinguish the scalar fields and we assume the F-term of inflaton dominates the energy density during inflation. 
Note that while $\delta K_1$ cannot be forbidden by any symmetries, some symmetries of the inflaton forbid $\delta K_2$.  
The resultant potential for ${\tilde L}_i, H_u$ and $H_d$ during inflationary era is 
\begin{align}
V({\tilde L}_i,H_u,H_d)=&\sum_{a={\tilde L}_i,H_u,H_d} -c_a H^2|\phi_a|^2 \notag \\
&+ \left | \frac{\lambda_{Lii}}{M} {\tilde L}_i^2 H_u + \frac{\lambda_H}{M} H_u  H_d^2\right |^2 + \left | \frac{\lambda_H}{M} H_u^2 H_d \right |^2 + \left | \frac{\lambda_{Lii}}{M} {\tilde L}_i  H_u^2\right |^2  \notag \\
& +\left [ \frac{\lambda_{Lii}}{2M} H a_L {\tilde L}_i^2 H_u^2 + \mathrm{h.c.} \right ] +\left [ \frac{\lambda_H}{2M} H a_H H_d^2 H_u^2 + \mathrm{h.c.} \right ] , \label{potential}
\end{align}
where $c_a$'s are real and $a_L$ and $a_H$ are complex parameters with their magnitude presumably of order of unity. 
Terms including $H$ arise due to the coupling of flat direction to the F-term of the inflaton $F_I$. 
If $\delta K_2$ is forbidden, the Hubble induced A-terms, 
\begin{equation}
\left [ \frac{\lambda_{Lii}}{2M} H a_L {\tilde L}_i^2 H_u^2 + \mathrm{h.c.} \right ] +\left [ \frac{\lambda_H}{2M} H a_H H_d^2 H_u^2 + \mathrm{h.c.} \right ] , 
\end{equation}
vanish. 
Hereafter, therefore, we consider the two cases with and without Hubble induced A-terms. 
In the remainder of this section and section \ref{Dyna}, we examine the case when the Hubble induced A-terms exist. 
In section \ref{noa}, we discuss the influence of existence or nonexistence of Hubble induced A-terms to the dynamics of the scalar fields and the resultant lepton asymmetry. 

Soft terms from hidden sector SUSY breaking, 
\begin{equation}
\sum_{a={\tilde L}_i,H_u,H_d} m_a^2|\phi_a|^2+\left [ \frac{\lambda_{Lii}}{2M} m_{3/2} A_L {\tilde L}_i^2 H_u^2 + \mathrm{h.c.} \right ] +\left [ \frac{\lambda_H}{2M} m_{3/2} A_H H_u^2 H_d^2 + \mathrm{h.c.} \right ] ,
\end{equation}
exist at the same time. 
Here $m_a$ and $m_{3/2}$ are masses of scalar fields and gravitino respectively, while $A_L$ and $A_H$ are complex numerical factors with their magnitude of order of unity. 
Scalar potential also includes D-terms, which are 
\begin{align}
V_D&=\frac{1}{2}(D_Y^2+D_1^2+D_2^2+D_3^2),  \\
D_Y&=\frac{g}{2}\left(\sum_i (|{\tilde L}_i^\uparrow|^2+|{\tilde L}_i^\downarrow|^2)-(|H_u^\uparrow|^2+|H_u^\downarrow|^2)+(|H_d^\uparrow|^2+|H_d^\downarrow|^2) \right) , \\
D_1&=\frac{g^\prime}{2}\left(\sum_i ({\tilde L}_i^{\uparrow *} {\tilde L}_i^\downarrow +{\tilde L}_i^{\downarrow *} {\tilde L}_i^\uparrow)+(H_u^{\uparrow *} H_u^\downarrow +H_u^{\downarrow *} H_u^\uparrow)+(H_d^{\uparrow *} H_d^\downarrow +H_d^{\downarrow *} H_d^\uparrow) \right) , \\
D_2&=\frac{g^\prime}{2}\left(\sum_i ({\tilde L}_i^{\uparrow *} {\tilde L}_i^\downarrow -{\tilde L}_i^{\downarrow *} {\tilde L}_i^\uparrow)+(H_u^{\uparrow *} H_u^\downarrow -H_u^{\downarrow *} H_u^\uparrow) +(H_d^{\uparrow *} H_d^\downarrow -H_d^{\downarrow *} H_d^\uparrow) \right), \\
D_3&=\frac{g^\prime}{2}\left(\sum_i (|{\tilde L}_i^\uparrow|^2-|{\tilde L}_i^\downarrow|^2)+(|H_u^\uparrow|^2-|H_u^\downarrow|^2)+(|H_d^\uparrow|^2-|H_d^\downarrow|^2) \right) , 
\end{align}
in this case. 
Here $g$ and $g^\prime$ are the gauge coupling constants and superscripts $\uparrow$ and $\downarrow$ are the indices of the representation of $SU(2)$. 
However, deviation from D-flat direction is strongly restricted. 
Therefore we restrict dynamics of scalar fields into the space of flat directions. 
Hereafter, we consider scalar fields parameterized as 
\begin {alignat}{5}
{\tilde L}_i&=
\begin{pmatrix}
0 \\ {\tilde \nu}_i
\end{pmatrix}, \ 
&H_u&=
\begin{pmatrix}
h_u \\ 0
\end{pmatrix}, \ 
&H_d&=
\begin{pmatrix}
0 \\ h_d
\end{pmatrix} ,\label{D-para}
\end{alignat}
and, 
\begin{equation}
\sum_i|{\tilde \nu}_i|^2-|h_u|^2+|h_d|^2=0 \label{D-const}. 
\end{equation}
We can easily see that $D_Y=D_1=D_2=D_3=0$ and thus $V_D=0$ by this parameterization and there remains no additional physical degrees of freedom. 
Here F-term ${\bf y_e}_{ij}|L_iH_d|^2$ in Ref.\cite{Senami:2002kn} vanishes because of D-flat conditions.  

Moreover, we assume 
\begin{equation}
\lambda_{L11}\sim \lambda_H \ll\lambda_{L22},\lambda_{L33} ,
\end{equation}
in other words, neutrino mass matrix has a hierarchy. 
In this case, potential minima is located at
\begin{align}
|{\tilde \nu}_1|&\sim (HM/\lambda_{L11})^{1/2} \\
|h_d|&\sim (HM/\lambda_H)^{1/2} \\
|h_u|^2&= |{\tilde \nu}_1|^2+|h_d|^2 \\
|{\tilde \nu}_2|&\sim |{\tilde \nu}_3|\simeq 0, 
\end{align} 
because ${\tilde \nu}_2$ and ${\tilde \nu}_3$ acquire heavy masses from $h_u$. 
Therefore only one slepton and two Higgs have large vevs and other scalar fields set in the origin approximately.  
Hereafter, we adopt a notation,  $\lambda_{L11}=\lambda_L$, ${\tilde \nu}_1={\tilde \nu}$ and $\lambda$ representing the mean value of $\lambda_L$ and $\lambda_H$.

\subsection{CP-violation}
We now comment on CP-violation. 
In the case only a single flat direction has a nonvanishing vev, there are two phase-dependent terms during inflationary era\cite{Dine:1995kz,Asaka:2000nb}, 
\begin{align}
\frac{aH\lambda \varphi^n}{nM^{n-3}}+{\rm h.c.} &= \frac{2|a|H\lambda |\varphi|^n}{nM^{n-3}}\cos(n\theta_\phi+\theta_a), \label{A-AD-hub} \\
\frac{Am_{3/2}\lambda \varphi^n}{nM^{n-3}}+{\rm h.c.} &= \frac{2|A|m_{3/2}\lambda |\varphi|^n}{nM^{n-3}}\cos(n\theta_\phi+\theta_A), \label{A-AD-ssb}
\end{align}
where $\varphi$ is the AD field parameterizing the flat direction,  $\theta_a$ and $\theta_A$ are phases of the coefficients $a$ and $A$, respectively. 
In this model, \eqref{A-AD-hub} determines the initial condition just after inflation and \eqref{A-AD-ssb} determines the eventual potential minimum in the angular direction for the following dynamics. 
Therefore $\theta_a-\theta_A$ represents the CP-violation. 
When $H \gg m_{3/2}$, in other words when \eqref{A-AD-ssb} is negligible, there is no CP violation, and lepton asymmetry cannot be induced because of Sakharov's condition \cite{Sakharov:1967dj}. 
As $H$ decreases to make the ratio of A-term from hidden sector to the total scalar potential for AD field larger, CP-violation becomes effective and lepton asymmetry can be generated.  
In our model, on the other hand, we have five phase-dependent terms during inflationary era, 
\begin{align}
 \frac{\lambda_{L}}{2M} H a_L {\tilde \nu}^2 h_u^2 + \mathrm{h.c.}  &= \frac{\lambda_L H}{M}|a_L||{\tilde \nu}|^2|h_u|^2\cos(2\theta_\nu+2\theta_u+\theta_l), \label{A-LHu-hub}\\
 \frac{\lambda_{L}}{2M} m_{3/2} A_L {\tilde \nu}^2 h_u^2 + \mathrm{h.c.}  &= \frac{\lambda_L m_{3/2}}{M}|A_L||{\tilde \nu}|^2|h_u|^2\cos(2\theta_\nu+2\theta_u+\theta_L),\label{A-LHu-ssb}\\
 \frac{\lambda_H}{2M} H a_H H_d^2 H_u^2 + \mathrm{h.c.}  &= \frac{\lambda_H H}{M}|a_H||h_u|^2|h_d|^2\cos(2\theta_u+2\theta_d+\theta_h) ,\label{A-HuHd-hub}\\
 \frac{\lambda_H}{2M} m_{3/2} A_H H_d^2 H_u^2 + \mathrm{h.c.}  &= \frac{\lambda_H m_{3/2}}{M}|A_H||h_u|^2|h_d|^2\cos(2\theta_u+2\theta_d+\theta_H), \label{A-HuHd-ssb}\\
 \left | \frac{\lambda_{L}}{M} {\tilde \nu}^2 h_u + \frac{\lambda_H}{M} h_d^2 h_u \right |^2 &\ni \frac{\lambda_L\lambda_H}{M^2}|{\tilde \nu}|^2|h_u|^2|h_d|^2 \cos(2\theta_\nu-2\theta_d), \label{F-phase}
\end{align}
where $\theta_\nu,\theta_u,\theta_d,\theta_l,\theta_L,\theta_h$, and $\theta_H$ are phases of ${\tilde \nu}, h_u, h_d, a_L, A_L, a_H$, and $A_H$, respectively. 
Four of them are A-terms and the last one is from F-term. 
In inflationary era, \eqref{A-LHu-hub}, \eqref{A-HuHd-hub} and \eqref{F-phase} are dominant and they determine the initial condition of scalar fields in the angular direction for the following dynamics. 
Moreover, in this case, the time dependence of \eqref{A-LHu-hub} and \eqref{A-HuHd-hub} change after inflation, on which we comment in the next section, therefore the potential minimum for angular direction is varied. 
In this sense, CP is violated strongly just after the end of inflation. 
In our model, therefore, lepton asymmetry can be induced just after the end of inflation. 

\subsection{Stability of multiple flat directions}
In the remainder of this section we demonstrate the necessity of considering multiple flat directions. 
When a flat direction acquires a large vev, many other flat directions acquire masses from F-term. 
Consequently such flat directions cannot take large values, so that single flat-direction description can be justified. 
However, there are directions that do not acquire F-term from other flat directions within the MSSM \cite{Gherghetta:1995dv}. 
The case of $LH_u$ and $H_uH_d$ is its typical example. 

The mass terms of scalar fields are 
\begin{equation}
-c_\nu H^2 |{\tilde \nu}|^2-c_u H^2 |h_u|^2 -c_d H^2 |h_d|^2, 
\end{equation}
neglecting soft terms from hidden sector. 
Setting the D-flat condition \eqref{D-const}, it reads
\begin{equation}
-(c_\nu+c_u) H^2 |{\tilde \nu}|^2 -(c_u+c_d) H^2 |h_d|^2.
\end{equation}
If $c_\nu+c_u>0$ and $c_u+c_d<0$, $h_d$ is stable at the origin and we conclude that $h_d$ falls down to the origin and we can take the one-flat direction description. 
However, if $c_\nu+c_u>0$ and $c_u+c_d>0$, both $\nu$ and $h_d$ are unstable and cannot stay at the origin. 
We can no longer take the description. 
Moreover, even if only $LH_u$ direction is selected in the beginning, $H_d$ is unstable at the origin for large parameter region. 
We can see this by evaluating mass matrix for $H_d$ around the local minimum of $LH_u$ flat direction parameterized as $\phi$ , 
\begin{align}
&-H^2
\begin{pmatrix}
c_d & 0 \\
0 & c_d 
\end{pmatrix}
+\left(\frac{\lambda_H}{M}\right)^2|\phi|^4
\begin{pmatrix}
1 & 0 \\
0 & 1 
\end{pmatrix} \notag \\
&+\frac{\lambda_{L}\lambda_H}{M}|\phi|^2
\begin{pmatrix}
(\mathrm{Re}(\phi))^2-(\mathrm{Im}(\phi))^2 & -2\mathrm{Re}(\phi)\mathrm {Im}(\phi) \\
-2\mathrm{Re}(\phi)\mathrm {Im}(\phi)  & -(\mathrm{Re}(\phi))^2+(\mathrm{Im}(\phi))^2
\end{pmatrix} \notag \\
&+\mathrm{Re}(a_H) \frac{\lambda_H H}{2M}
\begin{pmatrix}
(\mathrm{Re}(\phi))^2-(\mathrm{Im}(\phi))^2 & -2\mathrm{Re}(\phi)\mathrm {Im}(\phi) \\
-2\mathrm{Re}(\phi)\mathrm {Im}(\phi)  & -(\mathrm{Re}(\phi))^2+(\mathrm{Im}(\phi))^2
\end{pmatrix} \notag \\
&+\mathrm{Im}(a_H) \frac{\lambda_H H}{2M}
\begin{pmatrix}
-2\mathrm{Re}(\phi)\mathrm {Im}(\phi) & -(\mathrm{Re}(\phi))^2+(\mathrm{Im}(\phi))^2 \\
-(\mathrm{Re}(\phi))^2+(\mathrm{Im}(\phi))^2 & 2\mathrm{Re}(\phi)\mathrm {Im}(\phi)
\end{pmatrix}.\label{massmat}
\end{align}
The eigen values of this matrix are 
\begin{equation}
 \left(\frac{\lambda_H}{M}\right)^2|\phi|^4 -c_d H^2  \pm \left[\left(\frac{\lambda_{L}\lambda_H}{M^2}|\phi|^4+\mathrm{Re}(a_H) \frac{\lambda_H H}{2M}|\phi|^2\right)^2+\left(\mathrm{Im}(a_H) \frac{\lambda_H H}{2M}|\phi|^2\right)\right]^{1/2}. \label{eigen}
\end{equation}
As a consequence, $h_d$ is unstable around $LH_u$ flat direction with a wide range of parameters. 
For example, even when $c_d>0$, sufficiently large $a_H$ or $\lambda_H$ can make $h_d$ unstable. 
(Note, however,  that too large $\lambda_H$ makes ${\tilde \nu}$ stable.) 
Therefore, the situation where multiple flat directions have nonvanishing vevs is more natural than the case with single flat direction.

\section{Dynamics of scalar fields and leptogenesis\label{Dyna}}

In this section, we describe the motion of scalar fields and estimate baryon-to-entropy ratio. 
Here we consider only homogeneous mode and neglect fluctuations around it because the curvature of the potential is much larger than the Hubble parameter and quantum fluctuation is sufficiently suppressed in this case. 

In the Friedman universe with the spacetime metric, $ds^2=dt^2-a^2(t)dx^2$, scalar fields $\phi_a$ obey the equation of motion, 
\begin{equation}
\ddot{\phi_a}+3H\dot{\phi_a}+\frac{\partial V}{\partial \phi_a^*}=0.  \label{EOM}
\end{equation}
Parameterizing as \eqref{D-para}, there remains a D-term, 
\begin{equation}
V_D=(g^2+g^{\prime 2})(|\nu|^2-|h_u|^2+|h_d|^2)^2. \label{DD}
\end{equation}
Numerical analysis shows that this potential energy density remains small and that deviation from flat directions by this parameterization does not have great influence on the dynamics of scalar fields and the resultant lepton asymmetry. 
This result justifies our definition of the field variables \eqref{D-para} with which all the D-terms except \eqref{DD} vanish from the beginning. 
If $H^2 \gg \partial ^2 V(\phi)/\partial \phi_a \partial \phi_a^*$, the friction term is dominant and the fields are overdamped. 
On the other hand, if $H^2 \ll \partial ^2 V(\phi)/\partial \phi \partial \phi^*$, the inertial term is dominant and the fields are underdamped. 

In inflationary cosmology, the Hubble parameter evolves as follows. 
In inflationary epoch, the Hubble parameter is approximately constant and we call it $H_{inf}$. 
As inflation finishes, the inflaton starts oscillating around its minimum and the Hubble parameter decreases as in the  matter dominant era, $H=2/3t$.  
Meanwhile the inflaton decays into other light particles gradually, and finally the universe turns to radiation dominant. 
Note that the cosmic temperature reaches the highest just after the end of inflation as a result of partial decay of the inflaton even if preheating is not operative.  
As a consequence, we cannot neglect the effect of the high-temperature cosmic plasma on the scalar fields even during the inflaton oscillation dominant regime\cite{Anisimov:2000wx,Allahverdi:2000zd,Asaka:2000nb,Fujii:2001zr}. 

\subsection{Inflationary era}
\begin{figure}[htbp]
\begin{tabular}{c}
\begin{minipage}{1.0\hsize} 
\begin{center}
\includegraphics[width=10cm]{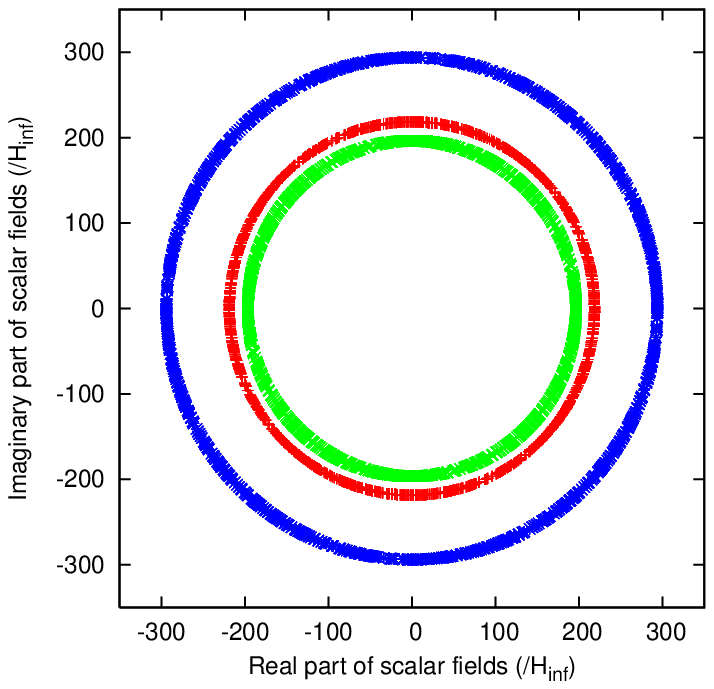}
\end{center}
\end{minipage}\\
\begin{minipage}{1.0\hsize}
\begin{center}
\includegraphics[width=10cm]{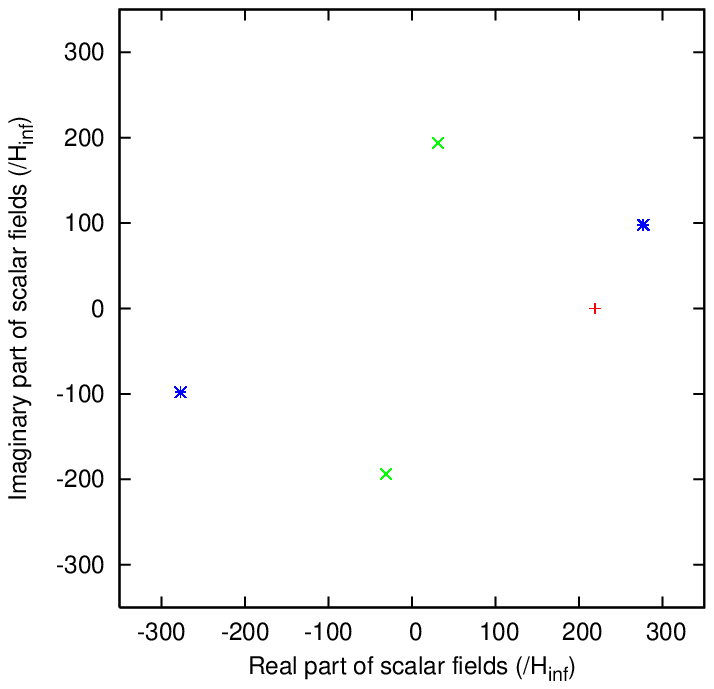}
\end{center}
\end{minipage}
\end{tabular}
\caption{The value of scalar fields at the end of inflation, the real part (horizontal axis) and the imaginary part (vertical axis) before (top) and after (bottom) gauge transformation. 
Red crosses represent ${\tilde \nu}$, blue x's represent $h_u$, and green stars represent $h_d$.   
The following parameters are used. $c_\nu=0.8$, $c_u=1.0$, $c_d=1.2$, $M/\lambda_L=1.2\times 10^5 H_{inf}$, $M/\lambda_H=1.5\times 10^5 H_{inf}$, $a_L=e^{i\pi/3}$, and $a_H=i$. 
Initial values are chosen randomly in the range $0.01H_{inf}\sim 100.0H_{inf}$.
\label{fig:one}}
\end{figure}

During inflationary epoch, scalar fields go to a minimum of the potential gradually. 
In our model, the potential \eqref{potential} is very complex and  it is nontrivial how many minima our model has and which minimum is selected by those fields. 
If the number of e-fold is large enough, scalar fields fall into one of the potential minima because $\partial^2 V(\phi)/\partial \phi \partial \phi^* \sim H_{inf}M/\lambda \gg H_{inf}^2$ at the minimum. 
Moreover, using numerical analysis we have confirmed that the minimum is unique except for gauge freedom and phase inversion. 
Figures \ref{fig:one} show typical example of the value where scalar fields fall from various initial values. 
Setting sneutrino real by $U(1)$ gauge transformation, $h_u$ and $h_d$ are found to be on only two minima whose amplitudes are equal and whose phase differences are $\pi$. 
What is important is that the values of the fields are fixed both in the angular direction and in the radial direction. 
In particular, CP-violation just after inflation arises from difference between the initial phases of scalar fields and the time-dependent potential minima in the angular direction at the inflaton oscillation dominant era. 
Note that if $\theta_l= \theta_h+\pi$, the potential minima during and after inflation are equivalent until the beginning of the oscillation of scalar fields around the origin. 
In this case, CP is not violated at large $H$. 
The change of potential minima is discussed precisely in the following section.

As a result, the vevs of scalar fields are fixed on the minima at the end of inflation, and following dynamics of scalar fields is independent of initial value at the beginning of inflation because quantum fluctuations are suppressed due to the large effective mass at the potential minima. 
In other words, there remains no pre-inflationary information. 
This feature is the same as in the one-field case. 

\subsection{Inflaton oscillation dominant era}
After inflation the universe enters the inflaton oscillation dominant era. 
The radial components of motion of scalar fields in this epoch are  almost the same as in the case of single flat direction. 

In this epoch the equation of motion is 
\begin{equation}
\ddot{\phi_a}+\frac{2}{t}\dot{\phi_a}+\frac{\partial V}{\partial \phi_a^*}=0. \label{EOM_mat}
\end{equation}
Neglecting phase dependent terms, and reparameterizing as 
\begin{align}
t&\rightarrow z=\ln(t/t_f),  &t_f&=\frac{2}{3}H_{inf}^{-1},\\
\phi_a&\rightarrow \chi_a=\phi_a/(HM/\lambda)^{1/2},
\end{align}
\eqref{EOM_mat} becomes\cite{Dine:1995kz} 
\begin{equation}
\frac{\partial^2 \chi_a}{\partial z^2}-\left( \frac{4}{9}c_a+\frac{1}{4} \right) \chi_a +\frac{4}{9}c_a\chi_a^5=0. 
\end{equation}
This equation has a fixed point $\bar{\phi_a}$\cite{Dine:1995kz}, 
\begin{equation}
\bar{\phi}_a \sim (10/9)^{1/4} \phi_a^{(0)}, \label{fix}
\end{equation}
where $\phi_a^{(0)}\sim(HM/\lambda)^{1/2}$ are minima of the potential. 
This fixed point presents a solution which traces points that are a little bigger than the time-dependent minimum of the potential. 
The initial values are the minima of the potential and different with these fixed points. 
Therefore scalar fields oscillate around the time-dependent fixed point with the period $\Delta z \sim 1$. 
Due to the time dependence of the potential minima, the oscillations do not redshift away even in the presence of cosmic expansion. 
As a result, scalar fields decrease with time as $\phi_a \sim (HM/\lambda)^{1/2} \sim (M/\lambda t) ^{1/2}$ oscillating around the fixed point. 

When a soft term from hidden sector, $m_{\phi_a}^2|\phi_a|^2$, 
or, thermal terms\cite{Anisimov:2000wx,Allahverdi:2000zd}, 
\begin{equation}
\sum_{f_k|\phi_a| <T}c_k f_k^2T^2|\phi_a|^2, \ \ \  \text{or} \ \ \ a_g \alpha_s(T)^2T^4 \ln \left(\frac{y_u^2|\phi_a|^2}{T^2}\right), 
\end{equation}
dominate the Hubble induced mass terms, the minima of the potential are at the origin and scalar fields begin oscillating around it. 
Here $k$ is an index of fields which couples to flat directions, $f_k$ is its coupling constant, $c_k$ and $a_g$ is numerical factor of order unity, and $\alpha_s(T)= g_s(T)^2/4\pi$. ($g_s$ is gauge coupling of $SU(3)$. )

When scalar fields oscillate around the origin, the amplitudes of scalar fields change with Hubble parameter as
\begin{equation}
|\phi_a| \propto 
\begin{cases}
H \ \ \text{(when driven by soft mass term from hidden sector)} \\
H^{7/8} \ \ \text{(when driven by thermal mass term)} \\
H^2 \ \ \text{(when driven by thermal log term)}.
\end{cases}\label{aftosc}
\end{equation}

\begin{figure}[htbp]
 \begin{center}
  \includegraphics[width=12cm]{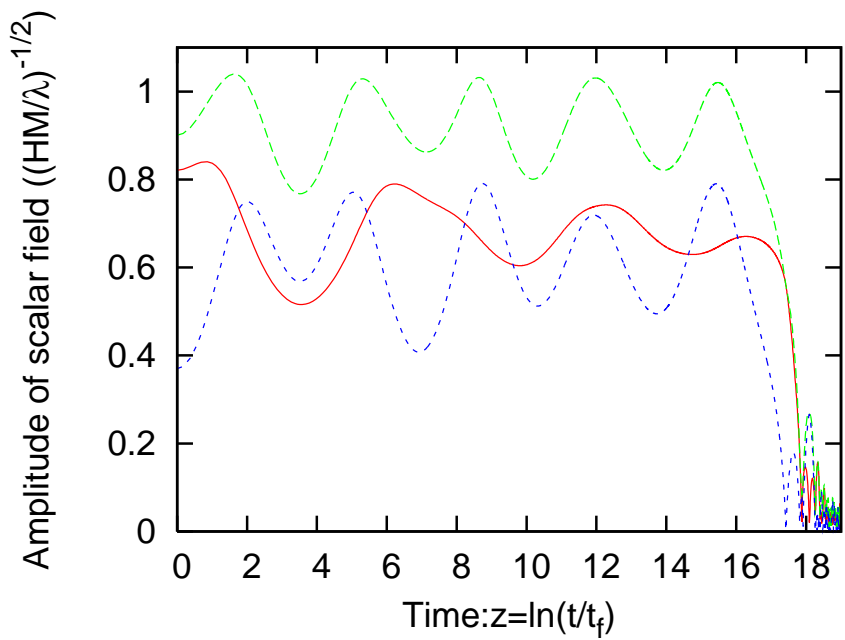}
 \end{center}
 \caption{Time variation of amplitudes of scalar fields. 
 Horizontal axis is logarithmic time and vertical axis is amplitude of scalar fields. 
Red solid line is ${\tilde \nu}$, green dashed line is $h_u$, blue dotted line is $h_d$. 
The following parameters are used. $c_\nu=c_u=c_d=1.0$, $H_{inf}=1.0 \times 10^{12}\mathrm{GeV}$, $M/\lambda_L=5.0\times 10^{20}\mathrm{GeV}$, $M/\lambda_H=7.5\times 10^{20}\mathrm{GeV}$, $a_L=e^{i\pi/6}$, $a_H=i$, and reheating temperature $T_R$ is $5.0\times 10^{7}$GeV.
 \label{fig:two}}
\end{figure}

Figure \ref{fig:two} shows typical time variation of scalar fields. 
The center of oscillation deviates from \eqref{fix} because the complicated mixings of scalar fields change the fixed points $\chi_a$. 

However, the angular motions  of scalar fields are different from one-field case. 
For this purpose, we first comment on the Hubble induced A-terms in this era. 
The Hubble induced terms in inflationary era, $-c_aH^2|\phi_a|^2$ and $(a\lambda H/M)\phi^n+{\rm h.c.}$, are from the couplings with $|F_I|^2$ and $F_I$, respectively. 
These couplings exist even in the inflaton oscillation dominant era. 
In this regime, inflaton oscillates around the potential minimum with sufficiently shorter time scale than the motions of AD fields. 
The induced potential, therefore, can be acquired by replacing $|F_I|^2$ and $F_I$ with their time average. 
Generally $\langle (\delta I)^2 \rangle$ is dominant in 
$|F_I|^2$ because the linear term $\langle \delta I \rangle$ vanishes by time averaging, where $\delta I$ is the fluctuation of inflaton from the potential minimum and $\langle \rangle$ represents time average. 
Moreover, when the F-term of inflaton includes second order monomials of the inflaton, $\langle |\delta I|^2 \rangle$ is also dominant in $| \langle F_I \rangle |$. 
As a consequence, Hubble induced A-terms have the same dependence on time as the Hubble induced mass terms. 
Since $\langle V_I \rangle=\langle |F_I|^2\rangle =3H^2M_G^2/2$ by the Virial theorem, the Hubble induced A-terms can be written in the form, 
\begin{equation}
V=\frac{a\lambda H^2}{\kappa H_{inf}M}I^4 + {\rm h.c.}, 
\end{equation}
where $\kappa$ is a numerical factor that is determined by the inflation model and here we assume that its magnitude ia order of unity. 
For example, if we take the superpotential of the form as a inflation model\cite{Yamaguchi:2000vm}, 
\begin{equation}
W(I,X)=v^2X(1-gI^2), 
\end{equation}
where $I$ is inflaton, and $X$ is auxiliary field, $v$ is vacuum energy,  $g$ is coupling constant whose dimension is $-2$, the important F-term is 
\begin{equation}
F_X=v^2(1-gI^2). 
\end{equation}
The resultant potential is 
\begin{equation}
V(I,X)=v^4[(1-gI^2)^2+4g^2I^2 X^2]
\end{equation}
In inflationary epoch, the vacuum energy dominates the universe, $V=V(0,X)=v^4$. Therefore, the Friedman equation reads  
\begin{equation}
v^4=3H_{inf}^2M_G^2
\end{equation}
After inflation, the inflaton oscillates around the potential minimum. 
As the potential minimum is 
\begin{equation}
I_0=\sqrt{1/g},
\end{equation}
we expand 
\begin{equation}
I=\sqrt{1/g}+\delta I
\end{equation}
Therefore, the F-term can be written as, 
\begin{align}
F_X&=-v^2[2\sqrt{g}\delta I+g(\delta I)^2], \\
|F_X|^2&\simeq 4gv^4|\delta I|^2,
\end{align}
and we take the time average of them, 
\begin{align}
|\langle F_X \rangle|&=v^2g\langle |\delta I|^2 \rangle, \\
\langle |F_X|^2 \rangle&=4gv^4\langle |\delta I|^2 \rangle. 
\end{align}
As a consequence, these two time averages are related as, 
\begin{equation}
|\langle F_X \rangle|=\langle |F_X|^2 \rangle/(4v^2)=\frac{3H^2M_G}{8\sqrt{3}H_{inf}}, 
\end{equation}
and we can write the Hubble induced A-term as, 
\begin{equation}
\langle F_X \rangle K^{X{\bar \phi}} D_\phi W\simeq \frac{a\lambda H^2}{H_{inf}M}\phi^4 + {\rm h.c.} \label{HAID}
\end{equation}
As the amplitude of AD field decreases in proportion to $H^{1/2}$, the Hubble induced A-terms fall off more rapidly than the mass terms. 
Note that the inflaton $X$ has $R$-charge 2 and the K${\ddot {\text a}}$hler potential of \eqref{NMK2} type break $R$-symmetry. 
Here we assume that $R$-symmetry of the inflaton is broken by some mechanism of high energy physics. 

\begin{figure}[htb]
\begin{tabular}{ccc}
\begin{minipage}{0.3\hsize} 
\begin{center}
\includegraphics[width=5cm]{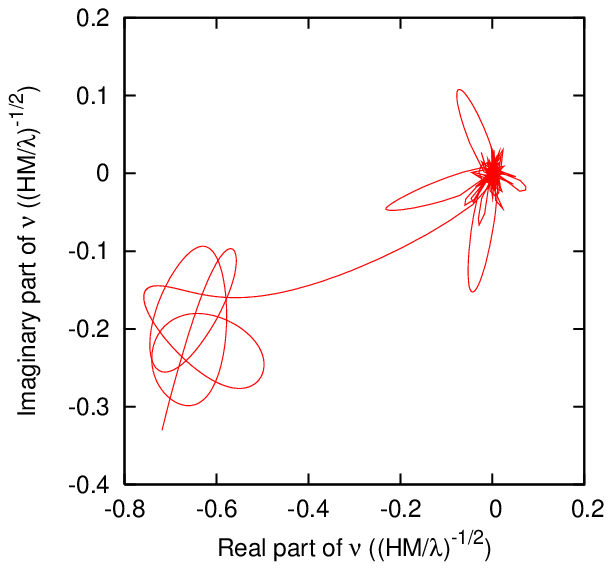}
\end{center}
\end{minipage}
\begin{minipage}{0.3\hsize}
\begin{center}
\includegraphics[width=5cm]{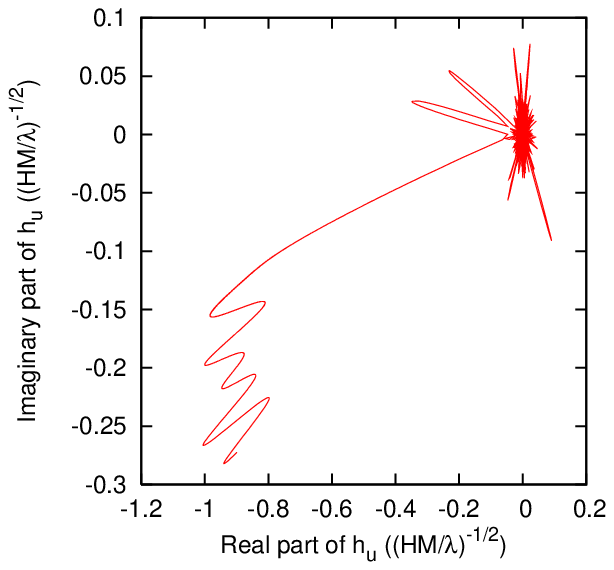}
\end{center}
\end{minipage}
\begin{minipage}{0.3\hsize}
\begin{center}
\includegraphics[width=5cm]{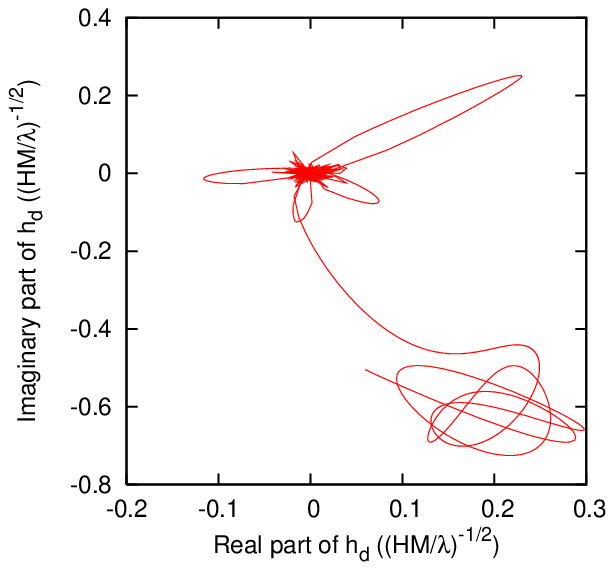}
\end{center}
\end{minipage}
\end{tabular}
\caption{Motion of scalar fields (${\tilde \nu}, h_u$, and $h_d$ from left to right) during inflaton oscillation dominant era. 
The value of fields is rescaled as $\chi=\phi/(HM/\lambda)^{1/2}$. 
The following value of parameters are used. $c_\nu=c_u=c_d=1.0$, $H_{inf}=1.0 \times 10^{12}\mathrm{GeV}$, $M/\lambda_L=1.0\times 10^{20}\mathrm{GeV}$, $M/\lambda_H=6.7\times 10^{19}\mathrm{GeV}$, $a_L=e^{i\pi/6}$, $a_H=i$, $\kappa=8$ and $T_R=5.0 \times 10^{7}$ GeV.
\label{fig:three}}
\end{figure}

In the one field case, the change of A-term discussed above does not make the change of the minima potential in the angular direction at large $H$ and it is fixed until just before the beginning of oscillation around the origin because the effect of CP violation is very small at that time. 
As we commented in the previous section, in our model, the phase difference of coupling constant of two A-term, \eqref{A-AD-hub} and \eqref{A-AD-ssb} yields CP violation. 
In contrast to the one field case,  there exist phase dependent terms in the potential \eqref{F-phase} and the Hubble induced A-terms. 
Moreover the change of A-terms \eqref{HAID} induces the change of potential minima in the angular direction. 
Therefore scalar fields also start oscillating around the minimum in the angular direction as in the radial direction at the onset of the inflaton oscillation. 
The equation of motion for the phase, $\theta_a$, of a scalar field $\phi_a$, 
\begin{equation}
{\ddot \theta_a}+\left(3H+\frac{1}{|\phi_a|^2}\frac{\partial}{\partial t}(|\phi_a|^2)\right){\dot \theta_a}+\frac{1}{|\phi_a|^2}\frac{\partial V}{\partial \theta_a}=0, 
\end{equation}
leads
\begin{equation}
\frac{\partial^2 \theta_a}{\partial z^2}+\frac{4}{9}\frac{H^{-2}}{|\phi_a|^2}\frac{\partial V}{\partial \theta_a}=0. 
\end{equation}
In this case, the relevant part of the potential is of the form of 
\begin{equation}
V=\frac{\lambda^2|\phi_a|^6}{M^2}\sin(2\theta_\nu-2\theta_d). \label{pot_ph}
\end{equation}
Here we approximate that all the scalar fields have the same amplitude. 
Therefore, the phases of ${\tilde \nu}$ and $h_d$ oscillate around its minimum without damping. 
The period of its motion is also $\Delta z = \CO(1)$. 
However \eqref{pot_ph} does not depend on the phase of $h_u$ and the only phase dependent terms, or the Hubble induced A-terms, for $h_u$ decreases rapidly and the potential for $h_u$ in the angular direction becomes flat. 
Therefore angular motion of $h_u$ is fixed immediately.
When the oscillation around the origin begins, the angular momentum is fixed to its value at that time because phase dependent term in the potential becomes negligible and it becomes the constant of motion. 
Figures \ref{fig:three} show typical motion of scalar fields. 
Scalar fields oscillate not only in the radial direction but also in the angular direction even before the oscillation around the origin begins. 
The angular motions of ${\tilde \nu}$ and $h_d$ are swinging, on the other hand, that of $h_u$ is not. 

\subsection{Lepton asymmetry}
Now we turn to the lepton asymmetry. 
Lepton-number density $n_L$ is expressed by the time component of the Noether current, or the operator, 
\begin{equation}
n_L(t)=i({\dot {\tilde \nu}}^*{\tilde \nu}-{\tilde \nu}^*{\dot {\tilde \nu}})=2|{\tilde \nu}|^2{\dot \theta_\nu}. 
\end{equation}
Therefore, using equation of motion we can derive the equation, 
\begin{equation}
{\dot n_L}+3H n_L=2\mathrm{Im}\left(\frac{\partial V}{\partial {\tilde \nu}}{\tilde \nu} \right). 
\end{equation}
In the integral representation, we find 
\begin{equation}
n_L(t)=\frac{2}{a(t)^3}\int_{t_f}^t dt^\prime a(t^\prime)^3 \mathrm{Im}\left(\frac{\partial V}{\partial {\tilde \nu}}{\tilde \nu} \right). 
\end{equation}
In the case of single flat direction, $LH_u$ \cite{Fujii:2001zr}, we evaluate the lepton asymmetry as 
\begin{equation}
n_L(t)=\frac{2}{a(t)^3} \int_{t_f}^t dt^\prime a(t^\prime)^3 \frac{\lambda_L m_{3/2}}{M} |\varphi|^4\sin (4\theta_\phi+\theta_a), 
\end{equation}
where $\varphi\equiv {\tilde \nu}=h_u$ represents AD field.
Here we consider only the hidden sector induced A-term and  neglect the Hubble induced A-term because AD field is located at its minimum in the angular direction and its contribution is negligible. 
The net lepton-to-entropy ratio is \cite{Fujii:2001zr}
\begin{equation}
\frac{n_L}{s}=\frac{T_R M}{12 \lambda_LM_G^2} \left( \frac{m_{3/2}}{H_{osc}} \right) \delta_{eff}. 
\end{equation} 
Here $\delta_{eff} \simeq \sin (4\theta_\phi+\theta_a)$ represents an effective CP violating phase. 
It has a factor $m_{3/2}/H_{osc}$ because the gradient of the potential to the angular direction is smaller than that to the radial direction by a factor of $m_{3/2}/H_{osc}$, 
which represents the dilution of CP violation. 
In the case of multiple flat directions, on the other hand, 
\begin{alignat}{3}
n_L(t)=&\frac{2}{a(t)^3}\int_{t_f}^t dt^\prime a(t^\prime)^3\left[ \frac{2\lambda_L}{\kappa M H_{inf}}H^2|A_L| |{\tilde \nu}|^2|h_u|^2\sin(2\theta_\nu+2\theta_u+\theta_L) \notag \right. \\
&\left.+ \frac{2\lambda_L}{M}m_{3/2}|A_L| |{\tilde \nu}|^2|h_u|^2\sin(2\theta_\nu+2\theta_u+\theta_L)+\frac{2\lambda_L\lambda_H}{M^2}|{\tilde \nu}|^2|h_u|^2|h_d|^2\sin(\theta_\nu-\theta_d)
\right] \notag \\
=& \frac{4}{a(t)^3}\int_{t_f}^t dt^\prime a(t^\prime)^3 \frac{\lambda}{M}H|a_L||\phi|^4\sin(\delta \theta) [1+\CO(\kappa^{-1})+\CO(m_{3/2}/H_{inf})] . \label{lepint1}
\end{alignat}
Here we set $t_f$ as the time when inflation finished 
We also use the fact that $|{\tilde \nu}|\simeq|h_u|\simeq|h_d|$ and $|\phi|$ represents the mean value of them. 

If sneutrino were in its phase minimum, $\mathrm{Im}[(\partial V/\partial {\tilde \nu}){\tilde \nu} ]=0 $, lepton asymmetry $n_L$ would not develop. 
In our model, however, sneutrino oscillates around the phase minimum during inflaton oscillation dominant era and does not stay at the minimum. 
The total lepton asymmetry can, then, be estimated as 
\begin{align}
n_\nu(t)&\simeq \frac{4}{a(t)^3}\int_{t_f}^{t} dt^\prime a(t^\prime)^3 |a_L|\frac{MH^3}{\lambda} \sin (\delta \theta) \notag \\
      &=\frac{8}{3}\left(\frac{a(t_f)}{a(t)}\right)^3 |a_L|\frac{MH_{inf}^2}{\lambda}\sin \Bigl(\delta_{eff}\left(t_{osc}\right) \Bigr) [1+\CO(H_{osc}/H_{inf})],\label{lepint4}
\end{align}
where $t_{osc}$ is the time when the oscillation around the origin starts and $\delta \theta \equiv 2\theta_\nu-2\theta_d$.  
Here $\delta_{eff}(t)$ is the total phase of the lepton asymmetry at $t$ that is due to the oscillation of $\delta \theta$.
For  $t>t_{osc}$, since the amplitudes of scalar fields decrease more rapidly than $H$, the contribution to the net lepton asymmetry in this epoch is suppressed by the factor $m_{3/2}/H_{inf}$. 
Here we neglect the Hubble induced A-term and the soft A-term because their contributions are very small. 
Moreover, we do not have a factor $\ln(t_{osc}/t_f)$ because of the time-dependence of $\delta \theta$. 

When reheating is finished, lepton-to-entropy ratio $n_L/s$ is fixed because both lepton number density and entropy decrease in proportion to $a^{-3}$ afterwards. 
The final lepton-to-entropy ratio is therefore given by 
\begin{align}
\frac{n_L}{s}(t)&=\frac{n_L}{s}(t_R) \notag \\
&=\frac{4}{(2\pi/45)g_{*s}T_R^3}\left(\frac{a(t_f)}{a(t_{R})}\right)^3|a_L|\frac{MH_{inf}^2}{\lambda}\sin \Bigl(\delta_{eff}(t_{osc}) \Bigr) \notag \\
&=\frac{4}{(2\pi/45)g_{*s}T_R^3}\left(\frac{(\pi^2 g_*/90)T_R^4}{3H_{inf}^2M_G^2}\right)^3|a_L|\frac{MH_{inf}^2}{\lambda}\sin \Bigl(\delta_{eff}(t_{osc})\Bigr) \notag \\
&=\frac{2|a_L|}{9}\left(\frac{g_*}{g_{*s}}\right)\frac{T_R M}{\lambda M_G^2}\sin \Bigl(\delta_{eff}(t_{osc}) \Bigr)
\end{align}
where $g_*$ is the total number of effectively massless degrees of freedom, $t_R$ is the time when reheating is finished and $M_G$ is reduced Planck mass.  
Without numerical factor of $\CO(1)$, we find, 
\begin{equation}
\frac{n_L}{s}(t)\simeq0.1 \times \frac{T_RM}{\lambda M_G^2} \sin \Bigl(\delta_{eff}(t_{osc})\Bigr). 
\end{equation}  
As a result, the sign of lepton-entropy ratio is determined by the information at $t=t_{osc}$ but its order is determined only by the cut-off scale $M$, the smallest coupling $\lambda$, and the reheating temperature $T_R$. 
Contrary to one-field description, there is no dilution factor $m_{3/2}/H_{osc}$, 
because in the present case CP is violated by the Hubble induced terms and is not weakened by thermal effect, which exists in the case of single flat direction. 
This implies that the net lepton asymmetry is determined almost by the parameters associated with the coupling between inflaton and flat direction and the low-energy SUSY breaking mechanism affects only a rather arbitrary phase factor $\delta_{eff}(t_{osc})$. 

While the number density of $h_d$ is created in the same mechanism, 
that of $h_u$ is rather small because it is created from the vanishing Hubble induced A-terms.    
This is the different feature with Ref. \cite{Senami:2002kn}. 

\begin{figure}[htbp]
 \begin{center}
  \includegraphics[width=12cm]{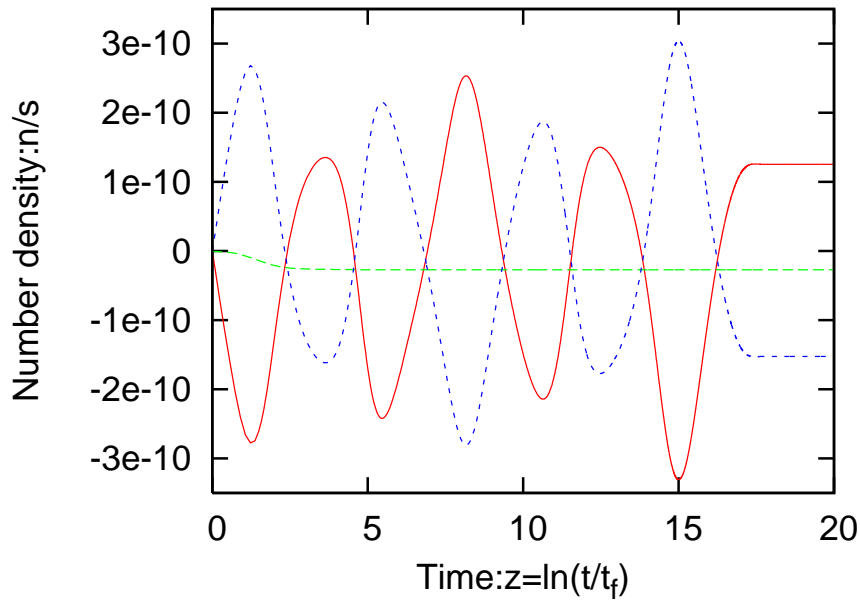}
 \end{center}
 \caption{Time evolution of lepton asymmetry and number density of Higgs fields par comoving volume. 
Horizontal axis is logarithmic time and vertical axis is lepton number density par comoving volume divided by entropy par comoving volume at the time when reheating is finished. 
Red solid line is lepton asymmetry, green dashed line is number density of $h_u$ and blue dotted line is that of $h_d$. 
The following parameters are used. $c_\nu=c_u=c_d=1.0$, $H_{inf}=1.0 \times 10^{13}\mathrm{GeV}$, $M/\lambda_L=1.0\times 10^{22}\mathrm{GeV}$, $M/\lambda_H=1.5\times 10^{22}\mathrm{GeV}$, $a_L=e^{i2\pi/3}$, $a_H=i$, and  $T_R=5.0\times 10^{6}$GeV.
 \label{fig:four}}
\end{figure}

\begin{figure}[htbp]
 \begin{center}
  \includegraphics[width=12cm]{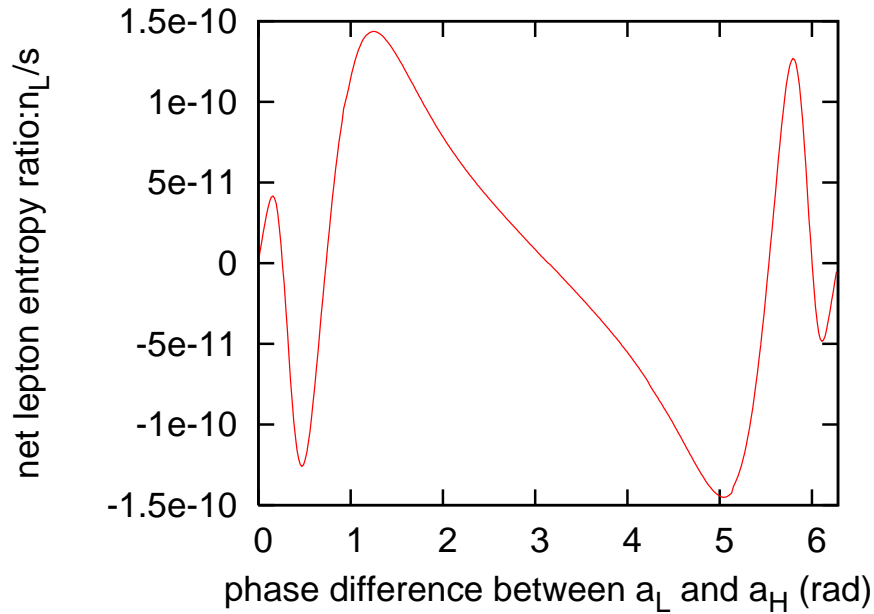}
 \end{center}
 \caption{The dependence of lepton asymmetry to phase difference between $a_L$ and $a_H$. 
Horizontal axis is phase difference and vertical axis is the net lepton-to-entropy ratio. 
The following parameters are used. $c_\nu=c_u=c_d=1.0$, $H_{inf}=1.0 \times 10^{13}\mathrm{GeV}$, $M/\lambda_L=1.0\times 10^{20}\mathrm{GeV}$, $M/\lambda_H=1.5\times 10^{20}\mathrm{GeV}$, and $T_R= 10^{8}$GeV.
 \label{fig:five}}
\end{figure}

Figure \ref{fig:four} shows typical time variation of the lepton asymmetry and the number density of Higgs fields. 
Lepton asymmetry and number density of $h_d$ par comoving volume fluctuates just after inflation finished, and fixed when oscillation around the origin begins. 
On the other hand, the number density of $h_u$ per comoving volume is fixed immediately after inflation because of the vanishing Hubble induced A-terms as discussed above. 
Figure \ref{fig:five} shows the net dependence of lepton asymmetry to phase difference between $a_L$ and $a_H$. 
When exactly $a_L-a_H=0$, or $\pi$, the net lepton asymmetry is equal to zero, as it should be. 

The sphaleron effect converts lepton asymmetry to baryon asymmetry as \cite{Khlebnikov:1988sr}
\begin{equation}
\frac{n_B}{s}=-\frac{8}{23}\frac{n_L}{s}. 
\end{equation}
Therefore present baryon-to-entropy ratio is represented as a function of reheating temperature and lightest neutrino mass, 
\begin{align}
\frac{n_B}{s}&\simeq -0.02\times \frac{T_RM}{\lambda M_G^2} \sin (\delta_{eff}(t_{osc})) \notag \\
&\simeq 10^{-10}\left(\frac{10^{-10}[\mathrm{eV}]}{m_\nu}\right)\left(\frac{T_R}{10^6[\mathrm{GeV}]}\right). \label{bary-ent}
\end{align}
Here we have taken the phase factor $\sin(\delta_{eff}(t_{osc})) = \CO(0.1)$ and used the equation \eqref{neutrinomass}  
assuming $\tan \beta = \CO(1)$.

\section{The case of no Hubble induced A-terms\label{noa}}

In the discussions so far, we have concentrated on the case with the Hubble induced A-terms. 
As mentioned in section \ref{sec-pot}, however, there is a case where Hubble induced A-terms is forbidden by some symmetry. 
In this section, we discuss the influence of no Hubble induced A-terms and comment on its cosmological implication. 

\subsection{CP-violation and stability of a flat direction}

In section \ref{sec-mod}, we have seen the CP-violation and stability of flat direction in our model. 
If the Hubble induced A-terms  \eqref{A-LHu-hub} and \eqref{A-HuHd-hub} are forbidden, the  situation is changed. 
During inflation only \eqref{F-phase} is the phase dependent term.  
The initial conditions of scalar fields in the angular direction for the following dynamics are not determined completely. 
The hidden sector induced A-terms \eqref{A-LHu-ssb} and \eqref{A-HuHd-ssb} are responsible for CP-violation. 
The degree of CP-violation has indefiniteness because of arbitrariness of initial phase. 
Furthermore, lepton asymmetry is induced at lower energy. 

As for the stability of a flat direction, we can set $a_L=a_H=0$ in the mass matrix \eqref{massmat} and the eigenvalues \eqref{eigen}. 
In \eqref{eigen} the large coupling constants of the Hubble induced A-terms are important for the instability of a flat direction. 
Therefore, the parameter space where we must consider multiple flat directions is smaller than in the case of existence of Hubble induced A-terms. 
Hereafter, however, we discuss the case where we must consider multiple flat directions. 

\subsection{Dynamics of scalar fields}

We turn to the difference of dynamics of scalar fields. 
While the dynamics in the radial direction is not changed, the dynamics in the angular direction is different. 

\subsubsection{Inflationary era}

During the inflationary era, the relevant potential for the angular direction is the cross coupling in the F-term, 
\begin{equation}
\frac{\lambda_L\lambda_H}{M^2}|\nu|^2|h_u|^2|h_d|^2 \cos(2\theta_\nu-2\theta_d) \in V. \label{FF}
\end{equation}
We can easily see the potential minimum for the phase of $\nu$ and $h_d$ is 
\begin{equation}
\theta_\nu-\theta_d=(2n+1)\pi \ \ \ (n\in \mathbb{Z})
\end{equation}
However, the potential for the phase of $h_u$ is flat and the initial phase of $h_u$ for the following dynamics is not determined by the potential. 
(Strictly speaking, considering the gauge freedom, the true physical degree of freedom is the phase difference between $h_u$ and $\nu$, $\theta_u-\theta_\nu$.) 
Therefore, the initial values for the following dynamics have indefiniteness. 
In other words, there remain pre-inflationary informations. 
In the following section, we see the influence of the indefiniteness for the resultant lepton asymmetry. 

\subsubsection{Inflaton oscillation dominant era}

\begin{figure}[htb]
\begin{tabular}{ccc}
\begin{minipage}{0.3\hsize} 
\begin{center}
\includegraphics[width=5cm]{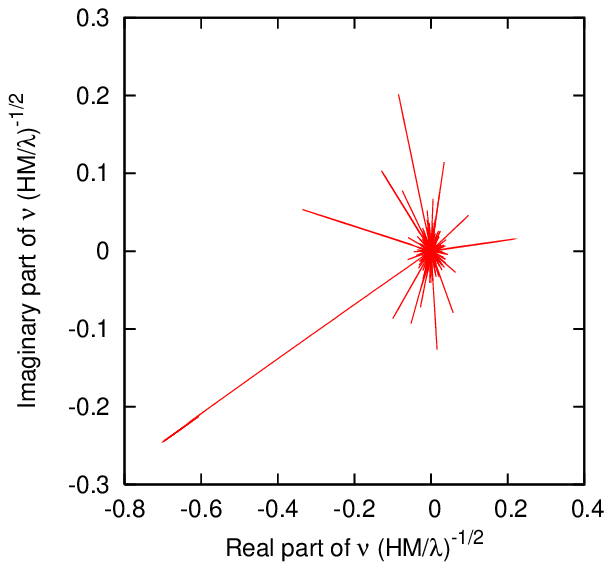}
\end{center}
\end{minipage}
\begin{minipage}{0.3\hsize}
\begin{center}
\includegraphics[width=5cm]{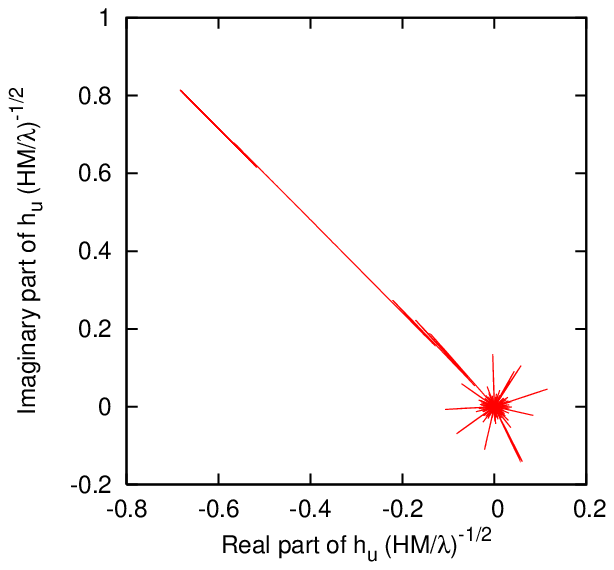}
\end{center}
\end{minipage}
\begin{minipage}{0.3\hsize}
\begin{center}
\includegraphics[width=5cm]{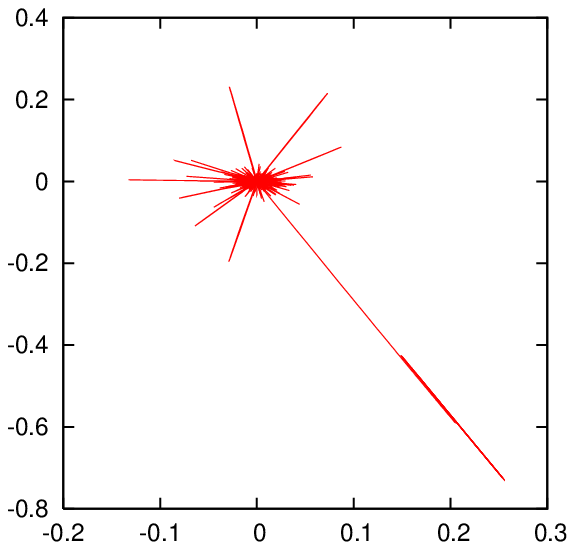}
\end{center}
\end{minipage}
\end{tabular}
\caption{Motion of scalar fields during inflaton oscillation dominant era in the complex plane, ${\tilde \nu}$, $h_u$, $h_d$ from left to right. 
The value of fields is rescaled as $\chi=\phi/(HM/\lambda)^{1/2}$. 
The following parameters are used. $c_\nu=c_u=c_d=1.0$, $H_{inf}=1.0 \times 10^{12}\mathrm{GeV}$, $M/\lambda_L=1.0\times 10^{20}\mathrm{GeV}$, $M/\lambda_H=3.3\times 10^{20}\mathrm{GeV}$, $A_L=e^{i\pi/6}$, $A_H=i$, and $T_R=10^{7}$GeV.
\label{fig:six}}
\end{figure}

We have seen that scalar fields begin to move in the angular direction just after the inflation when the Hubble induced A-terms exist. 
The source of this feature is the difference between the potential minima in the angular direction during inflation and at the inflaton oscillation dominant era. 
When the Hubble induced A-terms do not exist, however, the potential minima in the angular direction are not different between during and after inflation because \eqref{FF} is the only phase dependent potential for large $H$. 
In this case, therefore, scalar fields begin to move in the angular direction when the oscillations around the origin start as is the case of one field \cite{Dine:1995kz}. 
We can see this feature in Fig. \ref{fig:six}. 

\subsubsection{Lepton asymmetry}

\begin{figure}[htbp]
 \begin{center}
  \includegraphics[width=12cm]{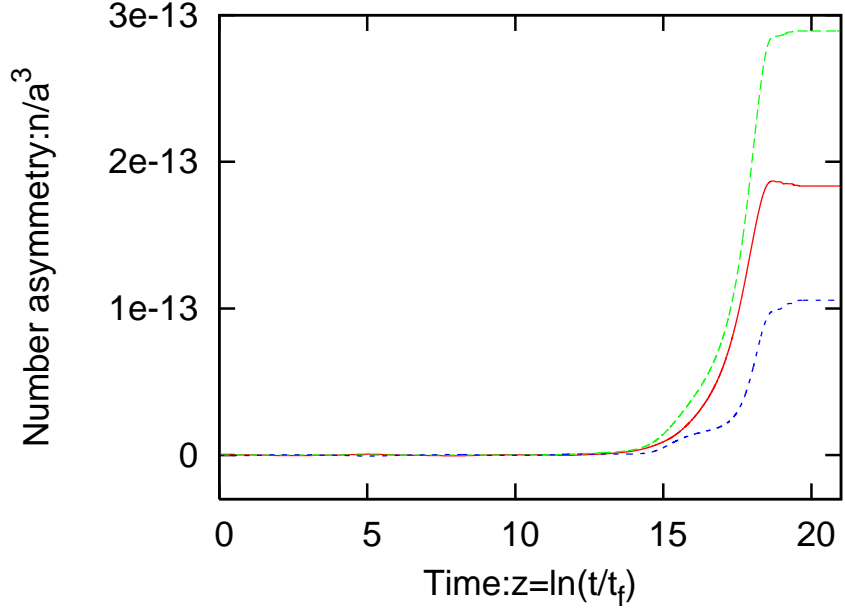}
 \end{center}
 \caption{Time evolution of lepton asymmetry and number density of Higgs fields par comoving volume. 
Horizontal axis is logarithmic time and vertical axis is lepton number density par comoving volume divided by entropy par comoving volume at the time when reheating is finished. 
Red solid line is lepton asymmetry, green dashed line is number density of $h_u$ and blue dotted line is that of $h_d$. 
The following parameters are used.  $c_\nu=c_u=c_d=1.0$, $H_{inf}=1.0 \times 10^{12}\mathrm{GeV}$, $M/\lambda_L=1.0\times 10^{20}\mathrm{GeV}$, $M/\lambda_H=3.3\times 10^{20}\mathrm{GeV}$, $A_L=1.0$, $A_H=i$, and  $T_R=1.0 \times 10^{7}$GeV.
 \label{fig:eight}}
\end{figure}

Now we comment on the resultant lepton asymmetry. 
We can estimate it by eq. \eqref{lepint1}. 
From this equation, we can see that the deviation from the potential minima in the angular direction is essential for generating lepton asymmetry. 
While deviation from potential minima exists just after the end of inflation when the Hubble induced A-terms exist, it does not exist until the beginning of the oscillation of scalar fields around the origin when the terms do not exist. 
In this case, moreover, the integrand is practically the same as that of one field case \cite{Dine:1995kz,Fujii:2001zr}. 
Therefore, the resultant lepton asymmetry is \cite{Fujii:2001zr}, 
\begin{equation}
\frac{n_L}{s}=\frac{MT_R}{12\lambda M_G^2}\left(\frac{m_{3/2}}{H_{osc}}\right)\delta_{eff}, 
\end{equation}
where $\delta_{eff}$ is the effective CP-violating phase. 
We can see that the suppression factor $m_{3/2}/H_{osc}$ is restored. 
Figure \ref{fig:eight} shows the typical time variation of the lepton asymmetry and the number density of Higgs fields. 
This feature is the same as one field case \cite{Dine:1995kz,Fujii:2001zr}. 

\begin{figure}[htbp]
 \begin{center}
  \includegraphics[width=12cm]{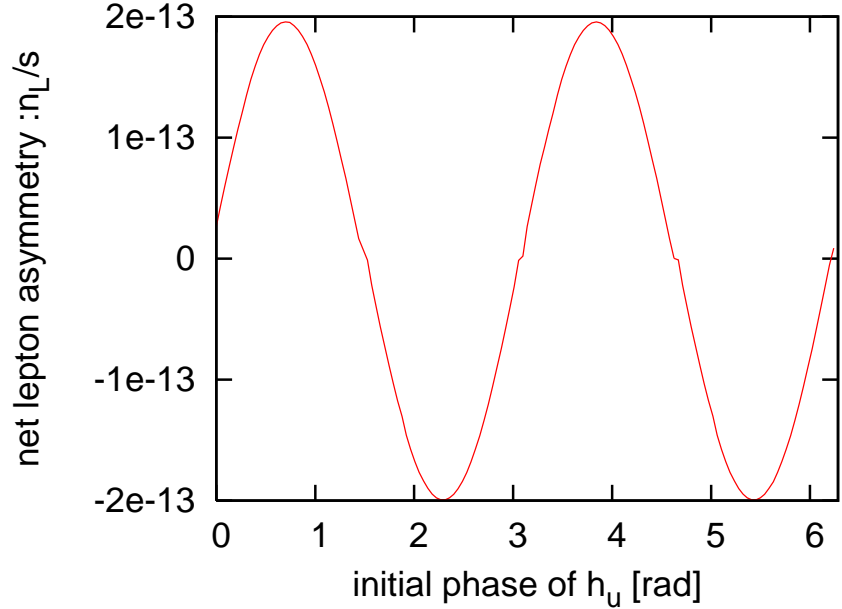}
 \end{center}
 \caption{The dependence of lepton asymmetry to the initial phase of $h_u$. 
Horizontal axis is the initial phase and vertical axis is net lepton entropy ratio. 
The following parameters are used. $c_\nu=c_u=c_d=1.0$, $H_{inf}=1.0 \times 10^{12}\mathrm{GeV}$, $M/\lambda_L=1.0\times 10^{20}\mathrm{GeV}$, $M/\lambda_H=3.3\times 10^{20}\mathrm{GeV}$, and $T_R = 10^{7}$GeV.
 \label{fig:seven}}
\end{figure}

The CP-violating phase $\delta_{eff}$ depends not only on the couplings of A-terms, $A_L$ and $A_H$, but also on the initial phase of $h_u$, which represents spontaneous CP-violation, as mentioned above. 
Figure \ref{fig:seven} shows the dependence on the initial phase of $h_u$.

\subsection{Baryon isocurvature perturbation}

Now we comment on baryon isocurvature perturbation\cite{Yokoyama:1993gb}. 
During inflation, in this case, the phase of $h_u$, $\theta_u$,  is effectively a massless field while other physical degrees of freedom have masses of order of $H$. 
Therefore, we must consider baryon isocurvature perturbation from the quantum fluctuation of $\theta_u$ as is the case with Ref. \cite{Senami:2007wz}. 

Assuming other degrees of freedom set at their potential minima, the equation of motion of $\theta_u$ is 
\begin{equation}
{\ddot \theta_u}+3H{\dot \theta_u}-\frac{1}{a^2}\Delta \theta_u=0, \label{eqan1}
\end{equation}
or we set $\varphi_u\equiv|h_u|\theta_u$, 
\begin{equation}
{\ddot \varphi_u}+3H{\dot \varphi_u}-\frac{1}{a^2}\Delta \varphi_u=0, \label{eqan2}
\end{equation}
Here we have taken the spatial derivative into account. 
Note that this equation of motion is applicable only if the fluctuation of $\theta_u$ is small enough. 
If we decompose the quantum fluctuation of $\varphi_u$, ${\hat \varphi_u}$,  as
\begin{equation}
{\hat \varphi_u(x)}=\int \frac{d^3 k}{(2\pi)^{3/2}}[a_k \varphi_{uk}(t) e^{i{\vect k \vect x}}+a_k^\dagger \varphi_{uk}^*(t) e^{-i{\vect k \vect x}}], 
\end{equation}
$\varphi_{uk}(t)$ satisfies the equation 
\begin{equation}
{\ddot \varphi_{uk}(t)}+3H{\dot \varphi_{uk}(t)}+{\vecs k}^2 a^{-2}\varphi_{uk}(t)=0\label{eqforph}
\end{equation} 
Here $a_k$ and $a_k^\dagger$ are annihilation and creation operators, respectively. 
In the de Sitter space, the solution of \eqref{eqforph} is 
\begin{equation}
\varphi_{uk}(t)=\frac{\sqrt{\pi}}{2}H_{inf}(-\eta)^{3/2}H_{3/2}^{(1)}(-k\eta), \ \ \eta=-(H_{inf}a(t))^{-1}, 
\end{equation}
where $H_{3/2}^{(1)}$ is the Hankel function of the first kind. 
Here the vacuum state is chosen so as to coincide with the usual Minkowski vacuum at $\eta \rightarrow -\infty$. 
Application of \eqref{eqan1} or \eqref{eqan2} is therefore justified since the fluctuation of $\theta_u$, ${\hat \theta_u} \sim H_{inf}/|h_u| \sim (H_{inf}\lambda/M)^{1/2} \ll 1$. 
 
Using the following expansion of the Hankel function, 
\begin{equation}
H_{3/2}^{(1)}(-k\eta)\simeq -\frac{i}{2\sqrt{\pi}}\left(\frac{-k\eta}{2}\right)^{-3/2},  \ \ \ \text{for}  \  -k\eta \ll 1
\end{equation}
we can see the power spectrum of the perturbation of $\varphi_u$, ${\mathcal P}(k)\equiv (k^3/2\pi^2)\langle \varphi_{uk}^*\varphi_{uk} \rangle$, is 
\begin{equation}
{\mathcal P}(k)=\left(\frac{H_{inf}}{2\pi}\right)^2. 
\end{equation}
The power spectrum of the fluctuation of the phase of $h_u$ is, therefore,  
\begin{equation}
{\mathcal P}_\theta(k)\equiv  \frac{k^3}{2\pi^2}\langle \theta_{uk}^*\theta_{uk} \rangle=\left(\frac{H_{inf}}{2\pi|h_u|}\right)^2 \simeq \left(\frac{\sqrt{\lambda H_{inf}}}{2\pi\sqrt{M}}\right)^2. 
\end{equation}
The fluctuation of the phase of $h_u$ generates the fluctuation of the net lepton asymmetry, $\delta n_L(k)$, which we can see from Fig. \ref{fig:seven}, 
\begin{equation}
n_L+\delta n_L(k) =  \frac{2}{3}\left(\frac{a(t_f)}{a(t)}\right)^3 |a_L|\frac{MH_{inf}^2}{\lambda}\sin (\delta_{eff}(t_{osc})+2{\hat \theta}_u(k)).
\end{equation}
Therefore baryon isocurvature perturbation is
\begin{equation}
\frac{\delta n_L(k)}{n_L}\simeq 2 \cot(\delta_{eff}(t_{osc})) {\hat \theta}_u(k).
\end{equation}
We can see from the observational constraint $n_B/s=0.9 \times 10^{-10}$ \cite{Spergel:2006hy} and \eqref{bary-ent}, 
\begin{equation}
\sin (\delta_{eff}(t_{osc}))\simeq -\frac{5\lambda M_G^2}{T_R M} \times 10^{-9}. \label{effpha}
\end{equation}
The power spectrum of baryon isocurvature perturbation is
\begin{align}
{\mathcal P}_B(k) &\equiv \frac{k^3}{2\pi^2} \left \langle \left(\frac{\delta n_L(k)}{n_L}\right)^2\right \rangle \notag \\
&=4\cot^2(\delta_{eff}(t_{osc})){\mathcal P}_\theta(k)=\left(\frac{\sqrt{\lambda H_{inf}}}{\pi\sqrt{M}}\right)^2\cot^2(\delta_{eff}(t_{osc})) \notag \\
&=  \left(\frac{\sqrt{\lambda H_{inf}}}{\pi\sqrt{M}}\right)^2 \left(\frac{4T_R^2M^2\times 10^{16}}{\lambda^2M_G^4}-1\right). 
\end{align}
Therefore larger reheating temperature and cut-off scale make larger isocurvature perturbation. 
Note that Eq. \eqref{effpha} means that $5\lambda M_G^2 \times 10^9 < T_R M$. 

The constraint on the baryon isocurvature perturbation in
terms of the ratio between the power spectrum of matter isocurvature perturbation ${\mathcal P}_{\mathcal S}(k)$
and that of curvature perturbation ${\mathcal P}_{\mathcal R}(k)$ is known as \cite{Kawasaki:2007mb} $B_a < 0.31$. 
Here  $B_a\equiv \sqrt{{\mathcal P}_{\mathcal S}(k)/{\mathcal P}_{\mathcal R}(k)}$. 
Using the observational data, ${\mathcal P}_{\mathcal R}=2.4\times10^{-9}$\cite{Spergel:2006hy}, and the relation, 
\begin{equation}
{\mathcal P}_{\mathcal S}(k)=\left(\frac{\Omega_b}{\Omega_m}\right)^2{\mathcal P}_B(k), 
\end{equation}
we find that parameters of this mechanism are constrained as 
\begin{equation}
\frac{\lambda H_{inf}}{M}\cot^2\Bigl(\delta_{eff}(t_{osc})\Bigr)=\frac{\lambda H_{inf}}{M}\times\left(\frac{4T_R^2M^2\times 10^{16}}{\lambda^2M_G^4}-1 \right)<7.0\times 10^{-8}. 
\end{equation}
Here $\Omega_b$ and $\Omega_m$ are the density parameters of baryon and matter, respectively, with their observed ratio $\Omega_b/\Omega_m=0.18$\cite{Spergel:2006hy}.  
This constraint requires small Hubble parameter during inflation. 
Moreover, small $\cot\Bigl(\delta_{eff}(t_{osc})\Bigr)$, or small reheating temperature or small cut-off scale is required. 

\section{Conclusion\label{conc}}

In this paper, we have obtained three important results about Affleck-Dine leptogenesis via multiple flat directions with non-renormalizable superpotential and vanishing renormalizable F-terms. 
First, when multiple flat directions have negative Hubble induced masses, we can no longer parameterize flat directions in terms of one complex scalar field and multi-dimensional motion of scalar fields must be considered. 
The scalar potential, however, has unique minimum except for gauge freedom and phase inversion if the Hubble induced A-terms exist. 
Therefore the degree of the mixing of flat directions is determined only by the shape of the potential without ambiguities, and the scalar-field dynamics in the post inflationary era are deterministic. 
Thus in the AD mechanism via multiple flat directions with nonrenormalizable potential to realize the nonvanishing field values, there remains no pre-inflationary information provided that inflation lasts long enough. 
In the case where the Hubble induced A-terms are absent, on the other hand, one of the physical phases of scalar fields takes an indefinitive value and its relative phase to the eventual potential valley acts as a source of CP-violation as in the case of single-field without the Hubble induced A-term. 

Second, there are CP-violation terms even for large $H$ due to cross coupling of scalar fields in non-renormalizable F-terms and the Hubble induced A-terms. 
Although the Hubble induced A-terms decreases rapidly after the end of inflation, 
they can give angular momentum to scalar fields. 
Therefore lepton asymmetry is generated just after the end of inflation. 
In particular, there is no suppression due to thermal effect \cite{Anisimov:2000wx}. 
Net lepton-to-entropy ratio does not have suppression factor $m_{3/2}/H_{osc}$, and the low-energy SUSY breaking mechanism, in particular, the gravitino mass, has practically no influence to the net lepton asymmetry. 
Since the Hubble induced A-terms play the critical role, the absence of them destroys this feature and the suppression factor  $m_{3/2}/H_{osc}$ is restored. 

Finally, in the case without the Hubble induced A-terms, this mechanism generates the baryon isocurvature perturbation and observational constraints narrow the allowed parameter space. 
In particular too large Hubble parameter during inflation is disfavored by this constraint\cite{Yokoyama:1993gb}.

\section*{Acknowledgments}
KK is grateful to  M.  Kawasaki, M.  Senami and T.T.  Yanagida for helpful discussions. 
This work was partially supported by JSPS Grant-in-Aid for Scientific Research Nos.16340076(JY), and 19340054(JY).

\end{document}